\documentclass[aps,prb,a4paper,superscriptaddress,amsmath,amssymb,twocolumn,showpacs]{revtex4} 
\usepackage{pdfpages}
\usepackage{graphicx,epsfig}

\allowdisplaybreaks
\begin{document}

\newcommand{\Avg}[1]{\langle \,#1\,\rangle_G}

\newcommand{\rr}{\mathcal{\hat{\rho}}}

\newcommand{\E}{\mathcal{E}}
\newcommand{\Lag}{\mathcal{L}}
\newcommand{\M}{\mathcal{M}}
\newcommand{\N}{\mathcal{N}}
\newcommand{\U}{\mathcal{U}}
\newcommand{\R}{\mathcal{R}}
\newcommand{\F}{\mathcal{F}}
\newcommand{\V}{\mathcal{V}}
\newcommand{\C}{\mathcal{C}}
\newcommand{\I}{\mathcal{I}}
\newcommand{\s}{\sigma}
\newcommand{\up}{\uparrow}
\newcommand{\dw}{\downarrow}
\newcommand{\h}{\hat{\mathcal{H}}}
\newcommand{\Hn}{\mathcal{H}}
\newcommand{\himp}{\hat{h}}
\newcommand{\g}{\mathcal{G}^{-1}_0}
\newcommand{\D}{\mathcal{D}}
\newcommand{\A}{\mathcal{A}}
\newcommand{\projR}{\hat{\mathcal{P}}_R}
\newcommand{\proj}{\hat{\mathcal{P}}_G}
\newcommand{\K}{\textbf{k}}
\newcommand{\Q}{\textbf{q}}
\newcommand{\hzero}{\hat{T}}
\newcommand{\io}{i\omega_n}
\newcommand{\eps}{\varepsilon}
\newcommand{\+}{\dag}
\newcommand{\su}{\uparrow}
\newcommand{\giu}{\downarrow}
\newcommand{\0}[1]{\textbf{#1}}
\newcommand{\ca}{c^{\phantom{\dagger}}}
\newcommand{\cc}{c^\dagger}
\newcommand{\aaa}{a^{\phantom{\dagger}}}
\newcommand{\aac}{a^\dagger}
\newcommand{\bba}{b^{\phantom{\dagger}}}
\newcommand{\bbc}{b^\dagger}
\newcommand{\da}{d^{\phantom{\dagger}}}
\newcommand{\dc}{d^\dagger}
\newcommand{\fa}{f^{\phantom{\dagger}}}
\newcommand{\fc}{f^\dagger}
\newcommand{\ha}{h^{\phantom{\dagger}}}
\newcommand{\hc}{h^\dagger}
\newcommand{\be}{\begin{equation}}
\newcommand{\ee}{\end{equation}}
\newcommand{\bea}{\begin{eqnarray}}
\newcommand{\eea}{\end{eqnarray}}
\newcommand{\ba}{\begin{eqnarray*}}
\newcommand{\ea}{\end{eqnarray*}}
\newcommand{\dagga}{{\phantom{\dagger}}}
\newcommand{\bR}{\mathbf{R}}
\newcommand{\bQ}{\mathbf{Q}}
\newcommand{\bq}{\mathbf{q}}
\newcommand{\bqp}{\mathbf{q'}}
\newcommand{\bk}{\mathbf{k}}
\newcommand{\bh}{\mathbf{h}}
\newcommand{\bkp}{\mathbf{k'}}
\newcommand{\bp}{\mathbf{p}}
\newcommand{\bL}{\mathbf{L}}
\newcommand{\bRp}{\mathbf{R'}}
\newcommand{\bx}{\mathbf{x}}
\newcommand{\by}{\mathbf{y}}
\newcommand{\bz}{\mathbf{z}}
\newcommand{\br}{\mathbf{r}}
\newcommand{\Ima}{{\Im m}}
\newcommand{\Rea}{{\Re e}}
\newcommand{\Pj}[2]{|#1\rangle\langle #2|}
\newcommand{\ket}[1]{\vert#1\rangle}
\newcommand{\bra}[1]{\langle#1\vert}
\newcommand{\setof}[1]{\left\{#1\right\}}
\newcommand{\fract}[2]{\frac{\displaystyle #1}{\displaystyle #2}}
\newcommand{\Av}[2]{\langle #1|\,#2\,|#1\rangle}
\newcommand{\av}[1]{\langle #1 \rangle}
\newcommand{\Mel}[3]{\langle #1|#2\,|#3\rangle}
\newcommand{\Avs}[1]{\langle \,#1\,\rangle_0}
\newcommand{\eqn}[1]{(\ref{#1})}
\newcommand{\Tr}{\mathrm{Tr}}

\newcommand{\Pg}{\mathcal{P}_G}

\newcommand{\Vb}{\bar{\mathcal{V}}}
\newcommand{\Vd}{\Delta\mathcal{V}}
\newcommand{\Pb}{\bar{P}_{02}}
\newcommand{\Pd}{\Delta P_{02}}
\newcommand{\tb}{\bar{\theta}_{02}}
\newcommand{\td}{\Delta \theta_{02}}
\newcommand{\Rb}{\bar{R}}
\newcommand{\Rd}{\Delta R}

\title{Finite-Temperature Gutzwiller Approximation from  Time-Dependent \\ Variational Principle}

\author{Nicola Lanat\`a}
\affiliation{Department of Physics and Astronomy, Rutgers University, Piscataway, New Jersey 08856-8019, USA}
\altaffiliation{Corresponding author: lanata@physics.rutgers.edu}
\author{Xiaoyu Deng}
\affiliation{Department of Physics and Astronomy, Rutgers University, Piscataway, New Jersey 08856-8019, USA}
\author{Gabriel Kotliar}
\affiliation{Department of Physics and Astronomy, Rutgers University, Piscataway, New Jersey 08856-8019, USA}

\date{\today} 

\begin{abstract}

We develop an extension of the Gutzwiller approximation to finite temperatures
based on the Dirac-Frenkel variational principle.
Our method does not rely on any entropy inequality, and is
substantially more accurate than the approaches proposed in previous works.
We apply our theory to the single-band Hubbard model at different fillings,
and show that our results compare quantitatively well with dynamical mean field theory
in the metallic phase.
We discuss potential applications of our technique within the framework of first principle
calculations. 
%
%

\end{abstract}

\maketitle


The Gutzwiller approximation (GA)~\cite{Gutzwiller1,Gutzwiller2,Gutzwiller3}
is a very useful tool 
in order to study the ground state of complex strongly correlated electron systems.
This important many-body technique
has been also formulated and implemented in
combination with density functional theory (DFT),~\cite{HohenbergandKohn}
e.g., in the LDA+GA approach,~\cite{Fang,Ho,Our-PRX}
which has been applied successfully to many
real materials.~\cite{PhysRevLett.110.096401,PhysRevLett.104.047002,PhysRevLett.108.036406,PhysRevLett.105.096401,julien,FeSe-FeTe,Our-Ce,Our-Ce-2,Our-PRX}
For strongly correlated metals, the accuracy of the GA
is comparable with dynamical mean field theory
(DMFT),~\cite{DMFT,dmft_book}
even though the GA is much less computationally demanding.
This property makes it an ideal theoretical tool,
as numerical speed is essential
for the purpose of studying and discovering new materials.

In order to study several temperature-dependent phenomena, such as 
structural and magnetic transitions and coherence-incoherence crossovers, 
it would be highly desiderable to have
at our disposal an extension to finite temperatures of the GA
as accurate as the ordinary theory for the ground state.
In fact, this would enable us to study these properties also for
correlated systems so complex to be out of the reach of the presently
available methods, such as DMFT.


An extension of the GA to finite temperatures has been
previously proposed in Refs.~\onlinecite{finiteT-GA-Wang,finiteT-GA-fabrizio}.
This approximation scheme is based on an exact entropy inequality
which enables to calculate an upper bound to the free energy,~\cite{finiteT-GA-fabrizio} 
and minimize it numerically.
Of course, underestimating the entropy using an entropy inequality --- rather
than calculating it exactly --- constitutes a source of approximation
not present in the ordinary
zero-temperature GA.
In particular, it has been shown that this additional source of approximation
generates a few pathologies of the theory, 
such as giving a negative entropy at low
temperatures.~\cite{finiteT-GA-Wang,finiteT-GA-fabrizio}

In this work we introduce an extension of the GA to
finite temperatures based on the Dirac-Frenkel variational
principle~\cite{dirac,frenkel,Dirac_Frenkel_mixed-states} and, in particular,
on the time-dependent GA theory~\cite{schiro,myTDGA} 
(that we generalize to mixed states).
Our method
does not rely on any entropy inequality, but
only on the variational principle and the Gutzwiller approximation
--- which are the same approximations done in the ordinary zero-temperature GA.
Consequently, as we are going to show, our theory improves considerably the method of
Refs.~\onlinecite{finiteT-GA-Wang,finiteT-GA-fabrizio},
and gives results in
good quantitative agreement with DMFT for correlated metals,
even though it is much less computationally demanding.

\emph{Imaginary-time evolution.---}
Let us consider a generic system of correlated electrons represented by a
Hamiltonian $\h$, and
define the imaginary-time evolution of a given initial
density matrix $\rr_0$ as follows:
\be
\rr(\tau) = e^{-\h \tau}\,\rr_0\,e^{-\h \tau}\,,
\label{solLag-intro}
\ee
i.e., according to the following differential equation:
\be
\partial_\tau\rr(\tau)=-(\h\rr(\tau)+\rr(\tau)\h)
\equiv -\{\h,\rr(\tau)\}\,.
\label{anticommutator}
\ee
Our aim consists in approximating the imaginary-time dynamics defined above
and use it to construct the state of $N$ electrons at temperature $T$.
In fact, if $\tau=\beta/2$ and $\rr_0=\hat{P}_N$ 
is the projector onto the subspace with $N$ electrons, 
Eq.~\eqref{solLag-intro} reduces to 
$\hat{P}_N\,e^{-\beta\h}$,
which represents a thermal state with $T\equiv 1/\beta$.~\cite{prl-dynamics-rho}

In order to derive our approximation scheme, it will be useful to think of
$\rr$ as the density matrix corresponding to an
ensemble of pure states $\{\ket{\Psi_n}\}$,
\be
\rr(\tau)\equiv\sum_n p_n\,\ket{\Psi_n(\tau)}\bra{\Psi_n(\tau)}\,,
\ee
where $p_n$ are fixed probabilities coefficients.
Within this definition, evolving $\rr$ according to Eq.~\eqref{solLag-intro}
amounts to evolve
all of the pure states of the ensemble according to the equation
\be
d\ket{\Psi_n(\tau)}=-\h\ket{\Psi_n(\tau)}\,d\tau 
\,.
\label{imag-S-evolution}
\ee
Note that Eq.~\eqref{imag-S-evolution}
resembles a Schr\"odinger evolution in imaginary time,
as it can be obtained from the ordinary real-time
Schr\"odinger evolution
\be
d\ket{\Psi_n(t)}=-i\h\ket{\Psi_n(t)}\,dt 
\label{real-S-evolution}
\ee
by substituting $dt\rightarrow -i\,d\tau$.

\emph{Real-time Dirac-Frenkel scheme.---}
Let us introduce the following action:~\cite{Dirac_Frenkel_mixed-states}
\bea
\mathcal{S}_{\{p_n\}}[\{\Psi_{n}(t)\}]&=&
\!\int_{t_i}^{t_f}\!dt\,\Lag_{\{p_n\}}[\{\Psi_{n}(t)\}]
\label{action}\\
\Lag_{\{p_n\}}[\{\Psi_{n}\}]&\equiv&\sum_n p_n\,
\Av{\Psi_n}{i\partial_t-\h}\,,
\label{action-Lag}
\eea
which depends parametrically on the probability coefficients
$p_n$ (that are fixed).
From now on we refer to Eq.~\eqref{action} as the Dirac-Frenkel action.
It can be readily verified that, regardless the values of $p_n$,
the exact solution of the
Lagrange equations for the ensemble of states $\{\ket{\Psi_n(t)}\}$ is
given by Eq.~\eqref{real-S-evolution}.

The key advantage of the Dirac-Frenkel
characterization of the time evolution outlined above is that
it allows us to build up a well-founded variational
approximation scheme for the time evolution [Eq.~\eqref{real-S-evolution}] as follows.

Let us assume that we want to solve approximately the time-dependent problem
by restricting the search of the solution
within a submanifold $\mathcal{M}$ of trial ensembles $\{\ket{\Psi_n}\}$.
%
Once we are able to evaluate the action $S$ along any given trajectory
in $\mathcal{M}$,
the Dirac-Frenkel variational principle provides us with a prescription
to approximate the instantaneous time evolution of 
any $\{\ket{\Psi_n}\}\in\mathcal{M}$.
Note that, by construction, this time evolution is such that 
$\{\ket{\Psi_n(t)}\}\in\mathcal{M}$ $\forall\, t$.

\emph{Application to the GA.---} For sake of simplicity, in this work the method will be formulated
for the single-band Hubbard model:
\be
\h = \sum_{k}\sum_{\sigma=\uparrow,\downarrow}\epsilon_k\,\cc_{k\sigma}\ca_{k\sigma}
+U\sum_R\cc_{R\uparrow}\ca_{R\uparrow}\cc_{R\downarrow}\ca_{R\downarrow}\,,
\label{hubb}
\ee
where $k$ is the momentum conjugate to the site label $R$ and $\sigma$
is the spin label.
The extension to multi-band Hubbard models is
straightforward, and its numerical implementation
will be discussed in a future work.
In order to benchmark our theory, we 
present finite-temperature calculations of the Hamiltonian
[Eq.~\eqref{hubb}] at different fillings $N/\mathcal{N} = 1+\delta$,
where $\mathcal{N}$ is the number of $k$-points and $\delta$
is the doping.

Here we want to search for
the saddle point of the Dirac-Frenkel
action within the manifold $\mathcal{M}_G$ of ensembles of Gutzwiller states
represented as follows:
\be
\{\ket{\Psi_n}\}=\{\proj\,\ket{\Psi_{0n}}\}\equiv\mathcal{M}_G\,,
\label{GA-ensemble}
\ee
where $\ket{\Psi_{0n}}$ are Slater determinants
and $\proj\equiv\prod_R\projR$ 
is an operator whose local components
are defined as
$\projR \equiv \sum_\Gamma \Lambda_\Gamma\,\ket{R,\Gamma}\bra{R,\Gamma}$,
where $\Lambda_\Gamma$ are numbers
and $\ket{R,\Gamma}\bra{R,\Gamma}$ are the projectors onto
the corresponding local many-body states
$\ket{R,\Gamma}\in\{\ket{0},\ket{R,\uparrow},\ket{R,\downarrow},\ket{R,\uparrow\downarrow}\}$.

The physical density matrix corresponding to the ensemble [Eq.~\eqref{GA-ensemble}]
is
$\rr_G\equiv\proj^\dagga\,\rr_0^*\,\proj^\dagger$,
where 
\be
\rr_0^*\equiv \sum_n p_n \,\ket{\Psi_{0n}}\bra{\Psi_{0n}} \,/\,
\sum_n p_n \,\langle\Psi_{0n}|\Psi_{0n}\rangle
\label{rho0}
\ee
is called variational density matrix. We assume that $\rr_0^*$ can be
represented as 
the Boltzmann distribution 
of a generic noninteracting Hamiltonian $\forall\,t$.
In order to
calculate the energy corresponding to $\rr_G$ 
--- which is necessary to evaluate the Dirac-Frenkel action, see Eq.~\eqref{action-Lag}, ---
the manifold of ensembles $\mathcal{M}_G$
is further
restricted by the so called Gutzwiller constraints:~\cite{fab,finiteT-GA-Wang,finiteT-GA-fabrizio}
\bea
\Tr[\rr^*_0\,{\hat{\mathcal{P}}^\dagger_{R}\hat{\mathcal{P}}^\dagga_{R}}]
\!\!&=&\!\!1\label{c1}\\
\Tr[\rr^*_0\,{\hat{\mathcal{P}}^\dagger_{R}\hat{\mathcal{P}}^\dagga_{R}
\,\cc_{R\sigma}\ca_{R\sigma}}]
\!\!&=&\!\!\Tr[\rr^*_0\,{\cc_{R\sigma}\ca_{R\sigma}}]
=[1+\delta]/2
\,.~~~\label{c2}
\eea
Furthermore, the GA is assumed, which 
is an approximation scheme that, 
as DMFT,~\cite{DMFT} becomes exact
in the limit of infinite coordination lattices.

As in Ref.~\onlinecite{lanata}, we introduce the matrix of slave-boson amplitudes:
\bea
\phi_{\Gamma\Gamma'} &=& \delta_{\Gamma\Gamma'}\Lambda_\Gamma\,\sqrt{P^0_\Gamma}
\label{defphi}\\
P^0_\Gamma&\equiv&\Tr\!\left[\rr^*_0\,\ket{R,\Gamma}\bra{R,\Gamma}\right]
\label{defP0}
\,.
\eea
Within the above definitions, the Gutzwiller constraints can be represented as:~\cite{lanata,finiteT-GA-fabrizio}
\bea
\Tr[\phi^\dagger\phi]
\!\!&=&\!\!1\label{c1phi}\\
\Tr[\phi^\dagger\phi\,F^\dagger_{\sigma}F^\dagga_{\sigma}]
\!\!&=&
\Tr[\rr^*_0\,{\cc_{R\sigma}\ca_{R\sigma}}]
=[1+\delta]/2
\,,~~\label{c2phi}
\eea
where
$[F_{\sigma}]_{\Gamma\Gamma'}\equiv\langle\Gamma\,|\ca_{R\sigma}|\,\Gamma'\rangle$.
Furthermore, it can be shown that $\phi\phi^\dagger$ represents the local reduced density
matrix in the basis $\{\ket{R,\Gamma}\}$, while the expectation values
of quadratic non-local observables is given by:
\be
\Tr[\rr_G\,\cc_{R\sigma}\ca_{R'\sigma}]= |\R|^2\, \Tr[\rr_0^*\,\cc_{R\sigma}\ca_{R'\sigma}]\,,
\ee
where
$\R=\Tr[\phi^\dagger F^\dagger_\sigma \phi F^\dagga_\sigma] / [1-\delta^2]^{-\frac{1}{2}}$.
%
Using the above equations, the GA Dirac-Frenkel Lagrange function
can be rewritten as follows:~\cite{Our-PRX}
\begin{widetext}
\bea
&&\Lag_{\{p_n\}}[\{\Psi_{0n}\};\phi,\R,\R^*,\D,\D^*]=
\sum_n p_n\,
\Av{\Psi_{0n}}{i\partial_t-|\R|^2\sum_{k}\sum_{\sigma=\uparrow,\downarrow}\epsilon_k\,\cc_{k\sigma}\ca_{k\sigma}}/\N
\\&&\;\;
+\Tr\!\left[\phi^\dagger i\partial_t\phi\right] -
\Tr\!\left[U\,\phi\phi^\dagger\,
F^\dagger_{\uparrow}F^\dagga_{\uparrow}F^\dagger_{\downarrow}F^\dagga_{\downarrow}\right] -
\sum_{\sigma=\uparrow,\downarrow}
\left(\Tr\!\left[\D\,\phi^\dagger F^\dagger_\sigma \phi F^\dagga_\sigma\right]
- \D\R\left[1-\delta^2\right]^{\frac{1}{2}}
+ \text{c.c.}\right)
\,.\label{Lagrange}
\nonumber
\eea
\end{widetext}
Note that, following Ref.~\onlinecite{Our-PRX}, we have
formally enforced the definition of $\R$ 
using the Lagrange multiplier $\D$.

The Lagrange equations
for the real-time dynamics induced by Eq.~\eqref{Lagrange} are
the following:
\bea
&&\left[i\partial_t-\h^\Re_{\text{qp}}[\R,\R^*]\right]
\,\ket{\Psi_{0n}}=0\;\;\forall\,n 
\label{d1real}
\\
&&\left[i\partial_t-H^\Re_{\text{emb}}[\D,\D^*]
\right]\,\phi = 0
\label{d2real}
\\
&&~~\R=\Tr\!\left[\phi^\dagger F^\dagger_\sigma \phi F^\dagga_\sigma\right]
\left[1-\delta^2\right]^{-\frac{1}{2}}
\\
&&~~\D=2\left[1-\delta^2\right]^{-\frac{1}{2}}\,
\Tr\!\left[\rho^*_0\frac{\partial}{\partial\R}\h^\Re_{\text{qp}}[\R,\R^*]\right]
\,,~~
\eea
where
\bea
\h^\Re_{\text{qp}}[\R,\R^*]\!&\equiv&\!
|\R|^2\sum_{k\sigma}\epsilon_k\,\cc_{k\sigma}\ca_{k\sigma}
\\
H^\Re_{\text{emb}}[\D,\D^*]\,\phi \!&\equiv&\!
\frac{\delta}{\delta \phi^\dagger}\!
\left\{\Tr\!
\left[U\,
\phi\phi^\dagger\,
F^\dagger_{\uparrow}F^\dagga_{\uparrow}F^\dagger_{\downarrow}F^\dagga_{\downarrow}
\right]\right.
\\
&+& 
\!\!\sum_{\sigma} \left.
\left(\Tr\!\left[\D\,\phi^\dagger F^\dagger_\sigma \phi F^\dagga_\sigma\right]+\text{c.c.}\right)
\right\}\phi
\,.~~~~
\eea
Note that the generator of the instantaneous evolution is quadratic
and identical for all of the $\ket{\Psi_{0n}}$,
and that also the evolution of $\phi$ resembles formally
a time-dependent Schr\"odinger equation.

The instantaneous real-time evolution described by the equations above
corresponds to apply well defined increments on all of the the
states of $\mathcal{M}_G$, see Eq.~\eqref{GA-ensemble}.
We may represent these increments as follows:
\be
d\ket{\Psi_n}=[(\partial_t\proj)\,\ket{\Psi_{0n}}
+\proj\,(\partial_t\ket{\Psi_{0n}})]\,dt\,.
\label{dpsi}
\ee

\emph{Imaginary-time dynamics.---}
Our goal consists in modifying the real-time GA dynamics
defined above in order to approximate the
imaginary-time evolution [Eq.~\eqref{imag-S-evolution}].

The formal similarity between Eqs.~\eqref{imag-S-evolution} and \eqref{real-S-evolution}
suggests us that it is possible to approximate the imaginary-time evolution of
$\{\ket{\Psi_{n}}\}$ 
simply by substituting $dt\rightarrow -i\,d\tau$
in Eq.~\eqref{dpsi}.
It can be readily verified that this prescription
would amount to update the
Gutzwiller variational parameters, see Eqs.~\eqref{defphi} and \eqref{defP0},
as follows:~\footnote{Note that for this system $P^0$ is constant, as it depends only
on the doping $\delta$, which is fixed.}
\bea
\left[\partial_\tau+\h^\Re_{\text{qp}}[\R,\R^*]\right]
\,\ket{\Psi_{0n}}&=&0\;\;\forall\,n 
\label{d01}
\\
\left[\partial_\tau+H^\Re_{\text{emb}}[\D,\D^*]
\right]\,\phi &=& 0\,.
\label{d02}
\eea
Unfortunately,
%
Eqs.~\eqref{d01} and \eqref{d02} violate
the Gutzwiller constraints,
see Eqs.~\eqref{c1phi} and \eqref{c2phi}.
Consequently, similarly to Ref.~\onlinecite{Vienna-PRL},
it is necessary to define a ``projection scheme'' in order
to enforce them at every time step.

Here we propose to enforce Eqs.~\eqref{c1phi} and \eqref{c2phi}
by using the following prescription:
\bea
\left[\partial_\tau+\h^\Im_{\text{qp}}[\R,\R^*,E_0]\right]
\,\ket{\Psi_{0n}}&=&0\;\;\forall\,n 
\label{d1}
\\
\left[\partial_\tau+H^\Im_{\text{emb}}[\D,\D^*,\lambda^c,E^c]
\right]\,\phi &=& 0\,,
\label{d2}
\eea
where the ``generators'' have been modified as follows:
\bea
\h^\Im_{\text{qp}}\!\!&\equiv&\!\!\h^\Re_{\text{qp}}-E_0
\\
H^\Im_{\text{emb}}\phi\!\!&\equiv&\!\!H^\Re_{\text{emb}}\phi \!+\!
\frac{\delta\Tr[\lambda^c\sum_{\sigma}\!\phi^\dagger\phi\,F^\dagger_\sigma F^\dagga_\sigma
\!-\!E^c\,\phi^\dagger\phi]}{\delta \phi^\dagger}
\,\phi
\,,~~~~
\eea
and $E_0(\tau)$ is constructed in order to enforce the normalization condition
of $\rho^*_0$, see Eq.~\eqref{rho0}, while $E^c(\tau)$ and $\lambda^c(\tau)$
are constructed in order to enforce Eqs~\eqref{c1phi} and \eqref{c2phi},
respectively.

We point out that the procedure defined above
enables us to recover the ordinary GA theory
for the ground state at $\tau\rightarrow\infty$.
In fact, within the formulation of Ref.~\onlinecite{Our-PRX}, the
GA parameters of the ground-state are obtained as
the ground states of $\h^\Im_{\text{qp}}$ and $H^\Im_{\text{emb}}$,
which correspond to a fix point of our imaginary-time dynamics.

It can be readily verified that Eq.~\eqref{d1}
implies that the imaginary-time evolution
of the variational density matrix is given by:
\bea
\rr^*_0(\tau)=P_N\,e^{-2\int_0^\tau d\tau'\left[Z(\tau')
\sum_{k\sigma}\epsilon_k\,\cc_{k\sigma}\ca_{k\sigma}
-E_0(\tau')\right]}
\label{rhostarzerostep}
\,,
\eea
where $Z(\tau')\equiv |\R(\tau')|^2$ is the Gutzwiller quasi-particle
weight, and $E_0(\tau')$ is constructed in order to enforce the normalization
condition of $\rr^*_0(\tau)$ for all imaginary
times.
In fact, Eq.~\eqref{rhostarzerostep} satisfies:
\be
\partial_\tau\rr^*_0(\tau)=-\{H^\Im_{\text{qp}}(\tau),\rr^*_0(\tau)\}\,,
\ee
which is consistent with Eq.~\eqref{d2}, and enables us to avoid
to keep track of the time evolution of all of the states of $\mathcal{M}_G$
(which would be practically impossible).

Note that, since we are in the thermodynamical limit, the expectation
values with respect to $\rr^*_0(\tau)$ can be evaluated in
the grand-canonical ensemble, i.e., we can assume that
\be
\rr^*_0(\tau)\propto
e^{-\beta^*_0(\tau)\left[
\sum_{k\sigma}\epsilon_k\,\cc_{k\sigma}\ca_{k\sigma}
-\mu^*_0(\tau)\hat{N}\right]}\,,
\label{rhostarzero}
\ee
where $\beta^*_0(\tau)\equiv 2\int_0^\tau d\tau'Z(\tau')$,
$\hat{N}$ is the number operator,
and $\mu^*_0(\tau)$ is such that the system has $N$ electrons
in average.

The imaginary-time evolution of the slave-boson amplitudes is
obtained by substituting Eq.~\eqref{rhostarzero} into the
Lagrange equations for $\phi,\lambda^c,\R,\D,E^c$ and solving
them numerically.

\begin{figure}
\begin{center}
\includegraphics[trim = 10mm 50mm 50mm 5mm, clip, width=8.3cm]{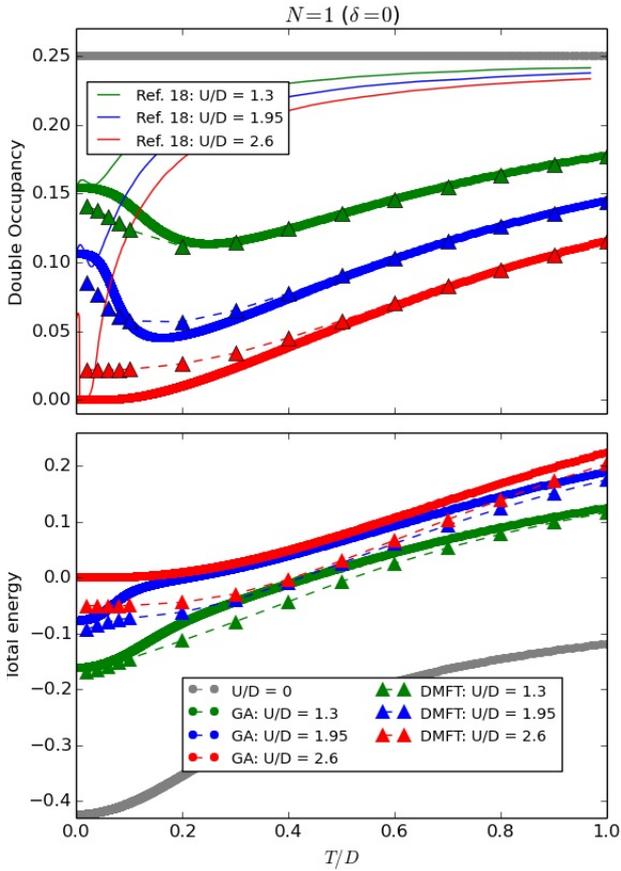}
\caption{GA calculations of the single-band Hubbard model
  at half-filling ($N=1$) in comparison with DMFT+CTQMC and the data of Ref.~\onlinecite{finiteT-GA-Wang}.
  Upper panel: evolution of the double occupancy as a function
  of the temperature. Lower panel: evolution of the total energy as a function
  of the temperature.
}
\label{figure1}
\end{center}
\end{figure}

\begin{figure}
\begin{center}
\includegraphics[trim = 10mm 50mm 50mm 5mm, clip, width=8.3cm]{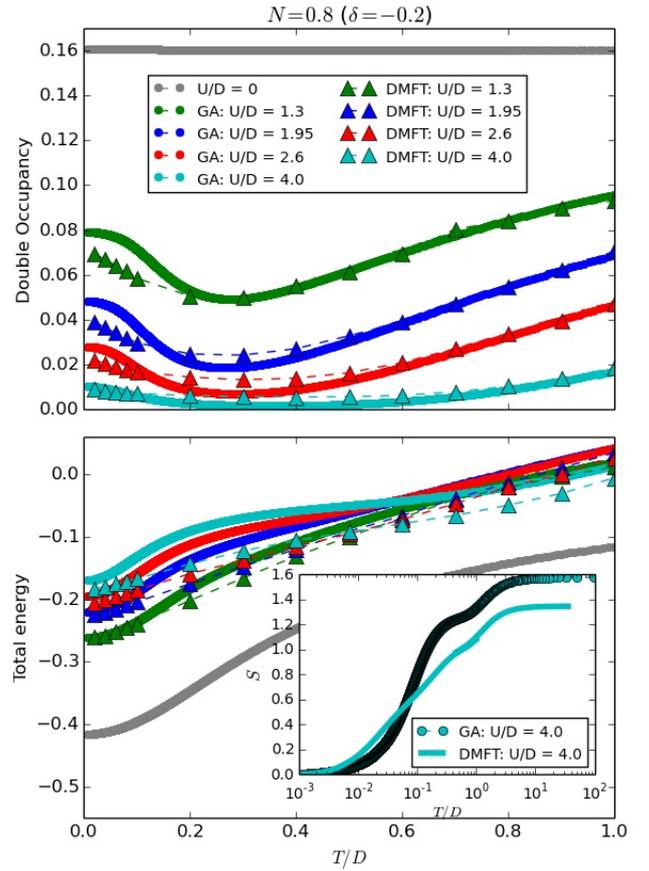}
\caption{GA calculations  of the single-band Hubbard model 
  away from half-filling ($N=0.8$) in comparison with DMFT+CTQMC.
  Upper panel: temperature dependence of the double occupancy.
  Lower panel: temperature dependence of the total energy $\E(T)$.
  Inset of the lower panel: temperature dependence (in logaritmic scale)
  of the GA entropy $S$
  in comparison with the DMFT data of Ref.~\onlinecite{Deng}.
}
\label{figure2}
\end{center}
\end{figure}

\emph{Numerical results.---}
Let us now discuss our numerical calculations of the
Hubbard model, see Eq.~\eqref{hubb}.
We 
assume a semicircular
density of states (corresponding to a
Bethe lattice in infinite dimensions)~\footnote{Note that DMFT is an exact theory for this system.}
and set the half-bandwidth $D$ as the unit of energy.
For comparison, we perform DMFT calculations
using the continuous time quantum Monte
Carlo method with hybridization expansion~\cite{ctqmc} as
impurity solver, as implemented in TRIQS.~\cite{TRIQS}

In the upper panel of
Fig.~\ref{figure1} is shown the evolution of the double occupancy
$d\equiv\langle\cc_{R\uparrow}\ca_{R\uparrow}\cc_{R\downarrow}\ca_{R\downarrow}\rangle$
as a function of the temperature at half-filling for several
values of $U$.
In the lower panel is shown the corresponding evolution of the
total energy $\E$.
The GA results are shown in comparison with DMFT and 
the Gutzwiller data of Ref.~\onlinecite{finiteT-GA-Wang}.

The agreement between the GA and DMFT+CTQMC is quantitatively
satisfying, especially for smaller values of $U$ and higher
temperatures (i.e., when the system is less correlated).
Indeed, our method improves substantially the results obtained
within the approximation scheme of Ref.~\onlinecite{finiteT-GA-Wang}.
The slight quantitative discrepancy for larger $U$'s
reflects the known fact that the Mott insulator is not well described
by the GA, but is approximated by the simple atomic limit
--- that is a state with $d=0$.
However, as long as the system is metallic,
our extension of the GA to finite temperatures is remarkably
accurate.

Let us now consider the Hubbard model away from half-filling.
In particular, we consider the case of $N=0.8$ electrons
per site (i.e., $\delta=-0.2$).
In the upper panel of
Fig.~\ref{figure2} is shown the temperature dependence
of the double occupancy for several values of $U$, while in
the lower panel is shown the evolution of the
total energy $\E$.
Finally, in the inset of the lower panel is shown the
temperature dependence of the entropy for $U/D=4$, in comparison
with the DMFT entropy calculated in Ref.~\onlinecite{Deng}.

We point out that, as discussed before, the entropy is not evaluated
directly from the GA variational parameters
(which could be done only approximately, e.g.,
by using the entropy inequality of Ref.~\onlinecite{finiteT-GA-fabrizio}),
but is calculated from the imaginary-time evolution of the total energy
using the well known thermodynamical identities $dS=d\E /T$, $S(T=0)=0$.
Note that the value of $S$ at $T\rightarrow\infty$ 
calculated according to these equations
depends on the evolution of the total energy
within the whole range of temperatures.
It is for this reason that the GA entropy shown in
Fig.~\ref{figure2} is slightly shifted with respect to DMFT at high temperatures
--- even though the atomic limit
belongs to the GA variational space, and is thus captured exactly by
our approximation
scheme.~\footnote{The reason why DMFT does not suffer this inconvenience is
that it is an exact theory in infinite dimensions (while the GA is a variational approximation).}

The agreement between the GA and DMFT+CTQMC
is even better for $N=0.8$ than for half-filling
(which is to be expected, as the doped system is metallic for all $U$'s).
In particular, it is remarkable that the agreement for $S$ is satisfying
for $U/D=4$, which is the largest interaction strength considered.

In conclusion, we have developed an extension of the Gutzwiller
approximation to finite temperatures
based on the Dirac-Frenkel variational principle.
Since our method does not rely on any entropy inequality, but
only on the variational principle and the Gutzwiller approximation,
it is as accurate as the ordinary GA theory for the ground state,
and improves substantially the method previously proposed in
Refs.~\onlinecite{finiteT-GA-Wang,finiteT-GA-fabrizio}.
We have performed benchmark calculations of the single-band
Hubbard model at different fillings, and compared our results
with DMFT+CTQMC, finding good quantitative agreement between
the two methods in the metallic phase.
We believe that our method will enable us to calculate
from first principles several important physical quantities
--- such as the specific heat,
the entropy and the temperature dependent structural properties ---
of strongly correlated systems presently too complex to be
studied with more accurate methods,
such as DMFT.

\begin{acknowledgments}

We thank Michele Fabrizio for useful discussions
and Qiang-Hua Wang for allowing us to use his data in Fig.~\ref{figure2}.
This work was supported by U.S. DOE Office of
Basic Energy Sciences under Grant No. DE-FG02-99ER45761 and
by NSF DMR-1308141.

\end{acknowledgments}

\bibliographystyle{apsrev}


\end{document}